\documentclass[twocolumn]{aastex62}

\newcommand{\Msun}{\mathrm{M_{\sun}}}

\def\gsim{ \lower .75ex \hbox{$\sim$} \llap{\raise .27ex \hbox{$>$}} }
\def\lsim{ \lower .75ex \hbox{$\sim$} \llap{\raise .27ex \hbox{$<$}} }

\def\jcap{JCAP}

\graphicspath{{./}{figures/}}

\received{April 6, 2020}
\revised{May 28, 2020}
\accepted{May 31, 2020}
\submitjournal{ApJ}

\shorttitle{Parametrizing the WDM mass function}
\shortauthors{Lovell et al.}

\begin{document}

\title{\bf Towards a general parametrization of the warm dark matter halo mass function}

\correspondingauthor{Mark R. Lovell}
\email{lovell@hi.is}

\author[0000-0001-5609-514X]{Mark R. Lovell}
\affil{University of Iceland \\
Dunhaga 5 \\
107 Reykjav\'ik, Iceland}

\begin{abstract}

Studies of flux anomalies statistics and perturbations in stellar streams have the potential to constrain models of warm dark matter (WDM), including sterile neutrinos. Producing these constraints requires a parametrization of the WDM mass function relative to that of the cold dark matter (CDM) equivalent. We use five WDM models with half-mode masses, $M_\mathrm{hm}=[1.3,35]\times10^{8}$~$\Msun$, spread across simulations of the Local Group, lensing ellipticals and the $z=2$ universe, to generate such a parametrization: we fit parameters to a functional form for the WDM-to-CDM halo mass function ratio, $n_\mathrm{WDM}(M_{X})/n_\mathrm{CDM}(M_{X})$, of ($1+(\alpha M_\mathrm{hm}/M_{X})^{\beta})^{\gamma}$. For $M_{X}\equiv$ virial mass of central halos we obtain $\alpha=2.3$, $\beta=0.8$, and $\gamma=-1.0$, and this fit is steeper than the extended Press-Schechter formalism predicts. For $M_{X}\equiv$ mass of subhalos we instead obtain $\alpha=4.2$, $\beta=2.5$ and $\gamma=-0.2$; in both mass definitions the scatter is $\sim20$~per~cent. The second fit typically underestimates the relative abundance of $z=2$ WDM subhaloes at the tens of per cent level. We caution that robust constraints will require bespoke simulations and a careful definition of halo mass, particularly for subhalos of mass $<10^{8}\Msun$.  

\end{abstract}


\section{Introduction} \label{sec:intro}

Recent observations of various astrophysical observables have opened up the possibility of measuring the dark matter halo mass function at redshifts $z<0.5$. Lensing flux-anomalies studies \citep{Hsueh19,Gilman19} and an analysis of gaps in stellar streams around the Milky Way (MW) \citep{Banik19} have claimed to detect large numbers of dark matter halos at the mass scale of $10^{8}-10^{9}$~$\Msun$. These analyses then offer the opportunity to place constraints on the presence, or otherwise, of a cut-off in the linear matter power spectrum. The result then provides information about which set of models the dark matter particle can be drawn from. If the dark matter particle belongs to the cold dark matter (CDM) family of candidates such as weakly interacting massive particles \citep[WIMPs;][]{Ellis_84} and QCD axions \citep{Turner91}, then there will be no cut-off at wavenumbers $k<100~h/$Mpc. If instead there is a cut-off in this wavenumber range, this would suggest the dark matter particle is from the warm dark matter (WDM) family of candidates, such as sterile neutrinos \citep{Dodelson94,Shi99,Laine08,Lovell16} or possibly gravitinos \citep{Pagels82}.  

The crucial step for comparing observational results to dark matter models is an accurate prediction for the halo mass function in each model. Parametrizing the CDM halo mass function is now a well established practice. In its simplest form it is set by the cosmological parameters and the statistics of the pre-galaxy formation epoch density field \citep[][]{Jenkins01,Sheth01,Ludlow16}, although baryon physics introduces very significant uncertainty, for example through the ejection of gas \citep{Despali17} or through destruction by the disk of the central galaxy \citep{GarrisonKimmel17}. The WDM case has proved much more challenging, for two reasons. First, WDM differs from CDM in that it constitutes a set of cosmologically distinct models, and the rich phenomenology of the sterile neutrino power spectra exhibits a wide variety of cut-off shapes and slopes \citep{Boyarsky09b,Schneider15,Lovell16}. Second, WDM $N-$body simulations suffer from the spurious fragmentation of filaments, which generates many more halos than the physical model would predict \citep{Wang07,Lovell14}. Methods have been developed to prevent this spurious fragmentation, including using the full six dimensional phase space information \citep{Angulo13} and adaptive softening \citep{Hobbs16}. However most CDM simulations are run with standard $N$-body solvers, and their WDM counterparts are run with the same code; therefore, methods have to be developed to identify and remove spurious subhaloes in postprocessing \citep{Lovell14}\footnote{ It has also been suggested that spurious numerical effects in $N$-body codes may lead to excessive stripping of subhalos \citep{vdBosch18}. This effect would apply to both WDM and CDM, so we assume that it is unlikely to be relevant for the subhalo mass function ratio.}.

The standard approach to the parametrization of the WDM halo mass function has been to multiply the CDM mass function by a WDM fitting function, $R_\mathrm{fit}$, as derived from comparing CDM and WDM simulations of the same volume, i.e.:

\begin{equation}
R_\mathrm{fit}=\frac{dN_\mathrm{WDM}}{d\log{M_{X}}}/\frac{dN_\mathrm{CDM}}{d\log{M_{X}}}\equiv n_\mathrm{WDM}/n_\mathrm{CDM},
\end{equation}

\noindent  where $N_\mathrm{WDM}$ and $N_\mathrm{CDM}$ are the number of WDM and CDM haloes respectively in each logarithmic mass bin for a given halo mass definition, $M_{X}$, and $n_\mathrm{WDM}$ and $n_\mathrm{CDM}$ are the differential halo mass functions; note that some studies instead use a more general $n_\mathrm{CDM}$ derived semi-analytically \citep{Giocoli10,Despali17}. These include the fit to halos in a cosmological box by \citet{Schneider2012}, and the fit to subhalos of a MW-analog halo reported by \citet{Lovell14}. Both of these early studies had drawbacks: \citet{Schneider2012} did not consider subhalos; \citet{Lovell14} instead used a series of zoomed simulations of just one MW-analog halo, such that it is not clear whether the results could be generalized across many halos, and both studies used WDM models with thermal relic masses $<2.3$~keV, which are in strong tension with Milky Way satellite counts \citep{Kennedy14,Lovell16} and Lyman-$\alpha$ forest analyses \citep[e.g.][]{Viel13,Baur16}. \citet{Lovell20} argued that such fits offer a poor approximation to extended Press-Schechter \citep[EPS][ ]{Press74,Bond91,Benson13} expectations for the mass function. Both fits used variations on the \citet{Bode01} and \citet{Viel05} thermal relic power spectra, which have different shapes to the particle physics-motivated sterile neutrino power spectra\footnote{\citet{Schneider15} applied a version of the \citet{Lovell14} algorithm to $z=5$ simulations of less extreme WDM models, and subsequently fit an EPS model, to which \citet{Benito20} then fit analytical functions.}. 

The best method currently available for making such comparisons to observations is to first specify the dark matter model parameters and observable of interest, and then perform $N$-body/hydrodynamical simulations of this system. This approach was taken for lensing arc analyses by \citet{Despali19b}, who used simulations of lensing halos at $z=0.2$ that featured two distinct sterile neutrino models plus the \citet{Lovell14} algorithm for removing spurious subhalos. Running high resolution simulations of this sort is computationally expensive, thus it is the exception rather than the rule.

In this study we instead use a wide variety of simulations of different environments and sterile neutrino models to derive approximations for $R_\mathrm{fit}$, with separate $R_\mathrm{fit}$ for centrals/isolated halos on the one hand and for subhalos on the other. The purpose of these fits is to provide a first order understanding of whether a given model is in tension with the observational data: we reiterate that the ability to reject any model with certainty requires bespoke simulations of the sort run by \citet{Despali19b}.    

This {\it Paper} is organised as follows. In Section~\ref{sec:sim} we present the simulations, in Section~\ref{sec:res} we present our derived halo mass functions, and in Section~\ref{sec:con} we draw our conclusions.

\section{Simulations} \label{sec:sim}

\begin{table*}
    \centering
      \caption{Table of simulations used in this study. The columns are: simulation set name, number of volumes used in each set, type of dark matter model, parent simulation suite,  galaxy formation model, mass of the simulation dark matter particle, final redshift of the simulation, cosmology from which the cosmological parameters were derived, half-mode mass of the dark matter model, and the references to the original papers, which are: (1) \citet{Lovell19a}, (2) \citet{Lovell20b}, (3) \citet{Lovell17b}, (4) \citet{Sawala16a}, (5) \citet{Despali19b} (6) \citet{Oppenheimer16} and (7) \citet{Schaye15}.}
    \begin{tabular}{lclccccccc}
        \hline
         \hline
        Simulation & $N_\mathrm{vols}$ & DM model & GF model & Suite & $m_\mathrm{DM}$ $[\mathrm{M}_{\odot}]$ & $ z $& Cosmology &
        $M_\mathrm{hm}[\mathrm{M}_{\odot}]$ & Original paper \\
        \hline
        AP-LA11 & 1 & LA11 & Ref & AP-HR & $5\times10^{4}$ & 0 & WMAP7 & $9.2\times10^{8}$ & (1) \\
        AP-LA9 & 1 & LA9 & Ref & AP-HR & $5\times10^{4}$ & 0 & WMAP7 & $2.6\times10^{8}$ & (2) \\
        AP-LA120 & 6 & LA120 & Ref & AP-MR & $6\times10^{5}$ & 0 & WMAP7 & $3.1\times10^{9}$ & (3) \\
        AP-LA10 & 6 & LA10 & Ref & AP-MR &  $6\times10^{5}$  & 0 & WMAP7 & $5.3\times10^{8}$ & (3) \\
        AP-CDM(HR) & 1 & CDM & Ref & AP-HR &  $5\times10^{4}$  & 0 & WMAP7 & N/A & (4) \\
        AP-CDM(MR) & 6 & CDM & Ref & AP-MR &  $6\times10^{5}$  & 0 & WMAP7 & N/A & (4) \\
 
        LH-LA11 & 4 & LA11 & Rec & Lens Halo & $2\times10^{6}$ & 0.2 & Planck &  $9.4\times10^{8}$ & (5) \\
        LH-LA8 & 4 & LA8 & Rec & Lens Halo & $2\times10^{6}$ & 0.2 & Planck & $1.3\times10^{8}$ & (5) \\
        LH-CDM & 4 & CDM & Rec & Lens Halo & $2\times10^{6}$ & 0.2 & Planck & N/A & (6) \\
   
        L25-LA120 & 1 & LA120 & Rec & 25~Mpc box & $1\times10^{6}$ & 2 & Planck & $3.5\times10^{9}$ & This work \\
        L25-CDM & 1 & CDM & Rec & 25~Mpc box & $1\times10^{6}$ & 2 & Planck & N/A & (7)
    \end{tabular}
    \label{tab:table1}
\end{table*}


The simulations used in this paper are derived from three simulation suites. We use the APOSTLE (which we label AP) WDM simulations presented in \citet{Lovell17b}, \citet{Lovell19a} and \citet{Lovell20b}; the lensing halo (LH) simulations of \citet{Despali19b}, and a previously unpublished simulation of a WDM 25~Mpc uniform resolution, cosmological box (L25) that was run to $z=2$. These simulations were derived from CDM counterparts: the AP simulations in \citet{Fattahi16} and \citet{Sawala16a}, the LH simulations from \citet{Oppenheimer16} and the L25 from \citet{Schaye15}.

All of the simulations were run with a variation of the EAGLE galaxy formation code \citep{Crain15,Schaye15}. This code is a heavily-modified version of the {\sc p-gadget3} code \citep{Springel08b}, and features cooling, star-formation \citep{Schaye08}, the growth of black holes \citep{Springel05,RosasGuevara15} and feedback from supernova and active galactic nuclei \citep{Booth09,DallaVecchia12}. The EAGLE model parameters were first calibrated for a simulation with a simulation dark matter particle mass of $9.7\times10^{6}\Msun$, and this calibration is known as the REFERENCE (`Ref') parameter set. The model was subsequently recalibrated for higher resolution simulations -- particle mass $1.0\times10^{6}\Msun$ -- and is therefore known as the RECALIBRATED (`Rec') parameter set. The LH and L25 simulations were performed with the Rec parameters, and the AP simulations were instead all run with the Ref parameters, despite having lower dark matter particle masses than both LH and L25.  

The halos were identified with the friends-of-friends (FoF) algorithm and decomposed into subhalos using the {\sc subfind} code \citep{Springel01a,Dolag09}. The largest {\sc subfind} halo in each FoF halo is defined as the central halo, and all other {\sc subfind} halos as subhalos / satellites. Spurious subhalos were flagged and removed with the Lagrangian region sphericity method of \citet{Lovell14}, using a sphericity cut $s=0.2$; note that we do not also apply their cut on halo mass. The AP simulations were performed with the WMAP7 cosmological parameters \citep{wmap11} whereas the other simulations instead used parameters derived from the first Planck results \citep{PlanckCP13}. The AP (LH) simulations are zoomed simulations, and therefore we only use halos/subhalos that are $<3$~pMpc ($<4$~pMpc) from the high resolution region's centre-of-potential, whereas for the L25 simulations we use halos from the entire box.

We use five sterile neutrino models in total. Each model is described by two parameters: a sterile neutrino mass and a lepton asymmetry, $L_6$; see \citet{Lovell16} for a discussion of these quantities. All five models use a sterile neutrino mass of 7.0~keV. The $L_6$ applied, in the order of most extreme (smallest wavenumber cut-off, or `warmest') to most like CDM, are: $L_6=120$ (labeled LA120), the warmest of any 7~keV sterile neutrino; $L_6=11.2$ (LA11), the warmest consistent with the sterile neutrino dark matter decay interpretation of the unexplained 3.55~keV X-ray line \citep{Boyarsky14a,Bulbul14}; $L_6=10$ (LA10), an intermediate case; $L_6=9$ (LA9), the coldest consistent with the 3.55~keV line; and finally $L_6=8$ (LA8) which the coldest of any 7~keV sterile neutrino. The matter power spectra of these five models are shown in fig.~1 of \citet{Lovell17b}. Note that the different values of $L_6$ result in changes to the resulting shape of the linear matter power spectrum as well as the location of the cut-off: the LA120 cut-off is the sharpest, and LA8 is the shallowest out of these lepton asymmetry choices.  

The LA120 model was applied in six medium resolution (labeled MR, dark matter particle mass $\sim6\times10^{5}$~$\Msun$) AP volumes and the L25 box; LA11 in one high resolution (HR, dark matter particle mass $5\times10^{4}$~$\Msun$) AP volume and four LH volumes; LA10 in six MR AP simulations; LA9 in a single HR AP volume and LA8 in four LH simulations. Based on the resolution study in the Aquarius simulations \citep{Springel08b}, we expect that the halo mass functions are resolved to better than 10~per~cent above the mass that corresponds to $\sim100$ particles. The important properties of all the simulations used in this paper are included in Table~\ref{tab:table1}.

\section{Results}
\label{sec:res}

The goal of our analysis is to derive a set of fitting functions to the ratio of the WDM and CDM mass functions, $R_\mathrm{fit}$. We adopt the functional form:

\begin{equation}
  R_\mathrm{fit} = n_\mathrm{WDM}/n_\mathrm{CDM} = (1+(\alpha M_\mathrm{hm}/M_\mathrm{x})^{\beta})^{\gamma}, 
  \label{eqn:fit}
\end{equation}
  
\noindent where $M_\mathrm{x}$ is the halo mass, for a given halo mass definition, and $\alpha$, $\beta$ and $\gamma$ are three fitting parameters. We chose this functional form to mirror the range of shapes of the input linear matter power spectrum \citep{Lovell16}: a characteristic cut-off scale ($\alpha$), an approximate asymptotic power-law slope ($\gamma\times\beta$) and also a shape parameter that governs how quickly the curve transitions from flat to the asymptotic slope ($\beta$); this form has also been used by \citet{Benito20}. $M_\mathrm{hm}$ is the half-mode mass, which is defined as the mass scale that corresponds to the power spectrum wavenumber at which the square root of the ratio of the WDM and CDM power spectra is 0.5. The value of $M_\mathrm{hm}$ for each simulation is given in Table~\ref{tab:table1}: note that because $M_\mathrm{hm}$ is sensitive to the cosmological parameters, it can vary by a few per~cent between simulations with the same DM model. 

We adopt two definitions of halo mass. For central halos we use the virial mass, $M_{200}$, defined as the mass enclosed within the co-moving radius of overdensity 200 times the critical density required for collapse. For subhalos we instead use the gravitational bound mass as determined by {\sc subfind}, $M_\mathrm{SUBs}$. We obtain separate fits -- i.e. separate sets of $\alpha$, $\beta$ and $\gamma$ values -- for the two mass definitions. 

We fit our parameters with the following algorithm. We compute the function in equation~\ref{eqn:fit} for a series of $\alpha$, $\beta$ and $\gamma$ parameter values and then calculate the logarithm of the ratio between this function and the measured $n_\mathrm{WDM}/n_\mathrm{CDM}$ functions at halo masses (both definitions) $10^{8}$, $10^{9}$, and $10^{10}$~$\Msun$ \footnote{We note that the $10^{8}$~$\Msun$ bin is only marginally resolved in the LH and LA25 simulations according to our 100 particle limit. The HR and MR AP simulations are resolved comfortably.}. We then calculate the sum of these ratios across all simulations, such that each DM model is given the same weight. We then select the fitting parameters that minimize this function: given that these fits are intended to be first order approximations we do not compute their $\chi^{2}$ values. Finally, we note that the number of haloes / subhaloes in our samples ranges from $\gsim70$ (AP-LA120, volume 1) to $>70000$ (L25-CDM).

\begin{figure*}
    \centering
    \includegraphics[scale=0.7]{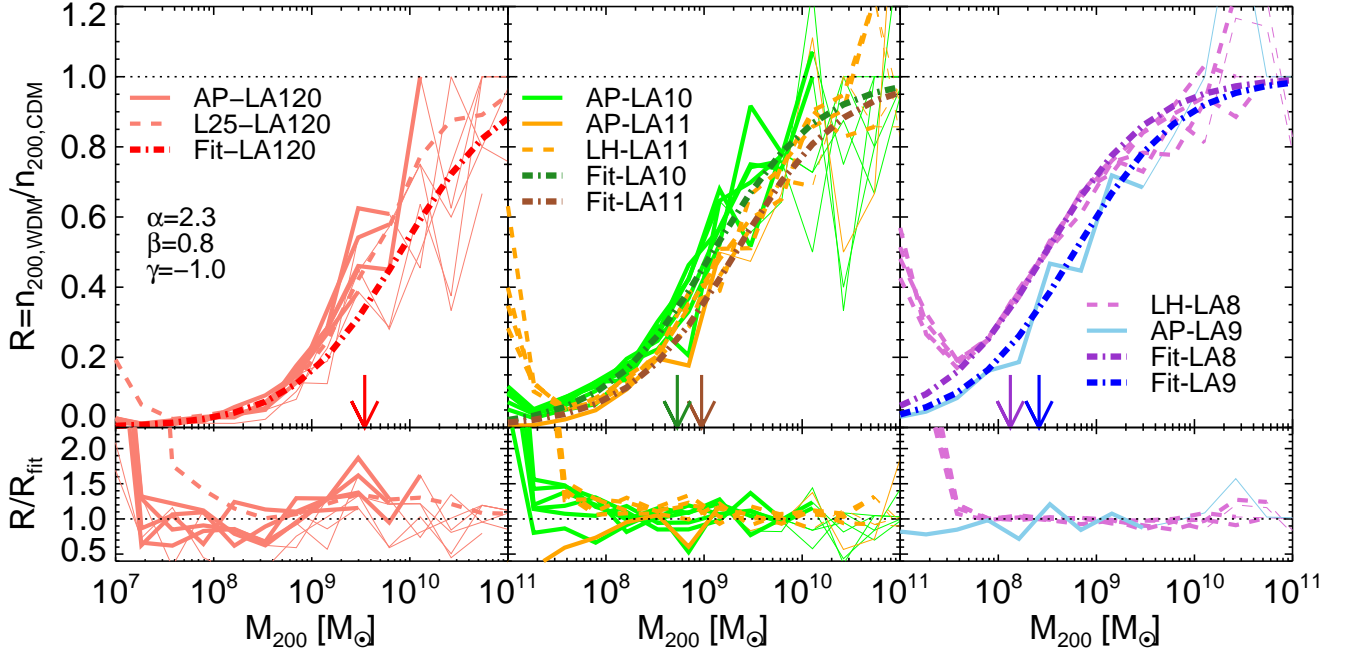}
    \caption{Ratio of WDM to CDM mass functions for central halos using the $M_{200}$ definition of mass. Top panels: the ratio of the WDM mass function to its CDM counterpart for models LA120 (red curves, left-hand panel), LA10 and LA11 (green and orange curves respectively, middle panel), and LA8 and LA9 (purple and blue curves respectively, right-hand panel). In all three panels, the solid curves indicate AP volumes. The dashed curves indicate the L25 box in the left-hand panel and the LH volumes in the other two panels. The fits are shown as dot-dashed curves, with a color darker than the corresponding model: the parameters of the fit are displayed in the left-hand panel. Bottom panels: the ratio of the data curves to the fit. In both sets of panels, curves are drawn thin where each mass bin contains fewer than ten WDM halos and are drawn thick elsewhere. The arrows mark $M_\mathrm{hm}$ for each of the models: where we use results from both Planck and WMAP7-derived simulations (LA120 and LA11) we plot the arrow using the Planck $M_\mathrm{hm}$ only.}
    \label{fig:m200}
\end{figure*}

We start our presentation of the resulting fits for the central halo mass function. We plot the ratios of $n_\mathrm{WDM}$ and $n_\mathrm{CDM}$ as a function of $M_{200}$ in Fig.~\ref{fig:m200}.

We obtain a fit that diverges by less than 50~per~cent in all of the mass bins $>10^{8}\Msun$ that contain more than ten halos. In lower mass bins, resolution effects in both WDM and CDM simulations are expressed. The parameters of this fit are: $\alpha=2.3$, $\beta=0.8$, and $\gamma=-1.0$, for an asymptotic low mass slope of $-0.8$. Remarkably, this slope almost exactly cancels the low mass slope determined for CDM subhalos -- not host halos -- in \citet{Despali17}, and therefore this fit predicts that the WDM halo mass function is approximately flat below the half-mode mass; a similar result was obtained in the study by the phase-space solver code of \citet{Angulo13}. We expect that a sharp cut-off in the halo mass function will occur at halo masses $<<M_\mathrm{hm}$, and determining the location of this cut-off will require a careful analysis of the delay in collapse time of low mass WDM haloes plus the statistics of the primordial density field; \citet{Benson13} argued that a much sharper cut-off will occur if thermal velocities are taken into account. 

\begin{figure*}
    \centering
    \includegraphics[scale=0.7]{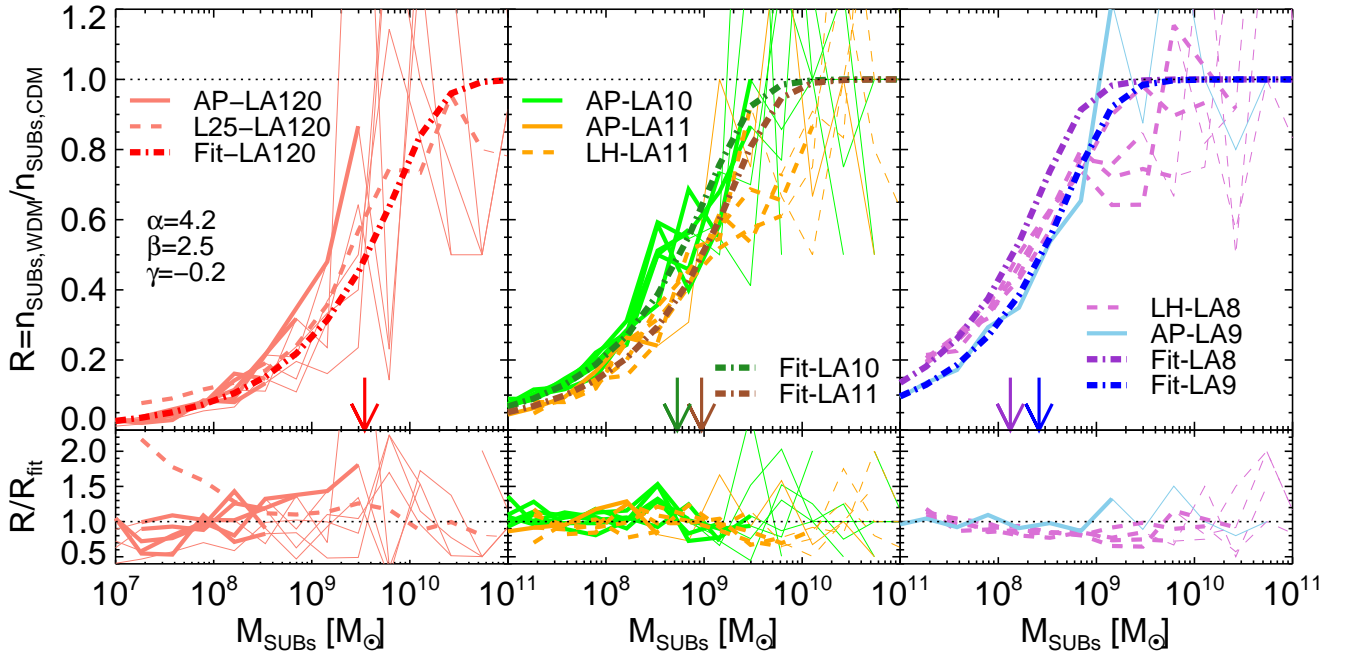}
    \caption{As for Fig.~\ref{fig:m200}, but for subhalos using the {\sc subfind} bound mass, $M_\mathrm{SUBs}$.}
    \label{fig:msubs}
\end{figure*}

Our fit works particularly well for the LA8 and LA9 models, producing agreement better than 20~per~cent, whereas LA120 is more challenging. There is a systematic bias for the fit to underestimate the number of LA120 halos at $\sim10^{9}\Msun$ and overestimate it at $\sim10^{8}\Msun$. This suggests that LA120 prefers a steeper fit than the other models, which may reflect the fact that the LA120 linear matter power spectrum cut-off is steeper than those of the other models. These fits are somewhat steeper than the EPS mass functions calculated by \citet{Lovell20}, which was calibrated using the high redshift conditional mass functions of $z=0$ centrals and is therefore a better match to the subhalo mass function than the isolated halo mass function. There are no conspicuous variations between fits attained in different environments, e.g. Local Groups vs. lensing halos. We repeat this exercise for the {\sc subfind} bound mass of satellite halos, $M_\mathrm{SUBs}$, in Fig.~\ref{fig:msubs}.

The parameters obtained for the satellite $M_\mathrm{SUBs}$ are $\alpha=4.2$, $\beta=2.5$ and $\gamma=-0.2$: the asymptotic slope is $-0.5$ and therefore shallower than for $M_{200}$; also, the turnover occurs at lower masses. This picture is consistent with the observation in \citet{Bose16a} that the difference between CDM and WDM increases for later forming halos, given that centrals form later than satellites. Another example of this is comparing the $z=2$ L25 simulation to the AP-LA120 volumes, where the former shows an excess of $10^{7}\Msun$ halos over the latter; we discuss the effect of halo formation time and redshift further in the appendix. In several instances the fit overestimates the abundance of $M_\mathrm{SUBs}<10^{8}\Msun$ halos by $\sim20$~per~cent, except for LA8 where instead it is $2\times10^{9}\Msun$ halos that are underestimated. We stress that this result relies on a particular definition of subhalo mass, and alternative subhalo finders may return very different results \citep{Onions12}. 

\section{Conclusions} \label{sec:con}

In this study we revisited the issue of parametrizing the suppression of the halo mass function in warm dark matter (WDM) relative to cold dark matter (CDM), taking advantage of hydrodynamical simulations at a variety of redshifts -- $z=0, 0.2$ and 2 -- and a variety of regimes -- the Local Group, lensing ellipticals and a uniform cosmological box -- to produce a mass function ratio fit, for the purpose of making first order comparisons between WDM models and observations. We find that the central halo differential mass function (using the virial mass definition, $M_{200}$) is approximated to typically better than 20~per~cent using equation~\ref{eqn:fit} with parameters $\alpha=2.3$, $\beta=0.8$, and $\gamma=-1.0$. This fit predicts a WDM halo mass function that is approximately flat for $M_{200}$ less than the half-mode mass, thus we expect to detect some $M_{200}\sim10^{6}$~$\Msun$ haloes in the WDM cosmology but at a drastically lower rate than in CDM. Performing a fit on the subhalo differential mass function (using the bound subhalo mass definition, $M_\mathrm{SUBs}$), with the same functional form, we obtain $\alpha=4.2$, $\beta=2.5$ and $\gamma=-0.2$, although the accuracy of this subhalo fit to the simulations is not as consistent as the $M_{200}$ fit. In particular, the $M_\mathrm{SUBs}$ fit consistently underpredicts the number of WDM haloes relative to CDM at $z=2$ by $\sim30$~per~cent: we hypothesise that this is due to the delay in collapse time for WDM haloes leading to greater disparities between CDM and WDM at late times \citep{Lovell19b}.

These fits are to be understood as first order approximation of each mass function. The simulations assume a particular model of galaxy formation and alternative implementations of feedback could return different results \citep{Despali17,GarrisonKimmel17}. Fitting the subhalo results is particularly challenging; future studies that invoke general fits such as those presented here will ultimately need to be backed up by bespoke, observables-oriented simulations, will need a clear definition of halo mass, and must take account of the differences between halo finders in order to claim definitive constraints on a given dark matter model.

\section*{Acknowledgments}

 MRL is grateful to the owners of the CDM simulations for their willingness to share their data, and would like to thank Wojtek Hellwing and Jes\'us Zavala for useful comments on the text. MRL acknowledges support by a Grant of Excellence from the Icelandic Research Fund (grant  number  173929). This work used the DiRAC@Durham facility managed by the Institute for Computational Cosmology on behalf of the STFC DiRAC HPC Facility
(www.dirac.ac.uk). The equipment was funded by BEIS capital funding
via STFC capital grants ST/K00042X/1, ST/P002293/1, ST/R002371/1 and
ST/S002502/1, Durham University and STFC operations grant
ST/R000832/1. DiRAC is part of the National e-Infrastructure. Part of this work was carried out on the Dutch National e-Infrastructure with the support of SURF Cooperative. This project has also benefited from numerical computations performed at 
the Interdisciplinary Centre for Mathematical and Computational Modelling (ICM) University of Warsaw 
under grants \#no GA67-15, GA67-16 and  G63-3

\section*{Appendix}
\label{sec:app}

\begin{figure*}
    \centering
    \includegraphics[scale=0.7]{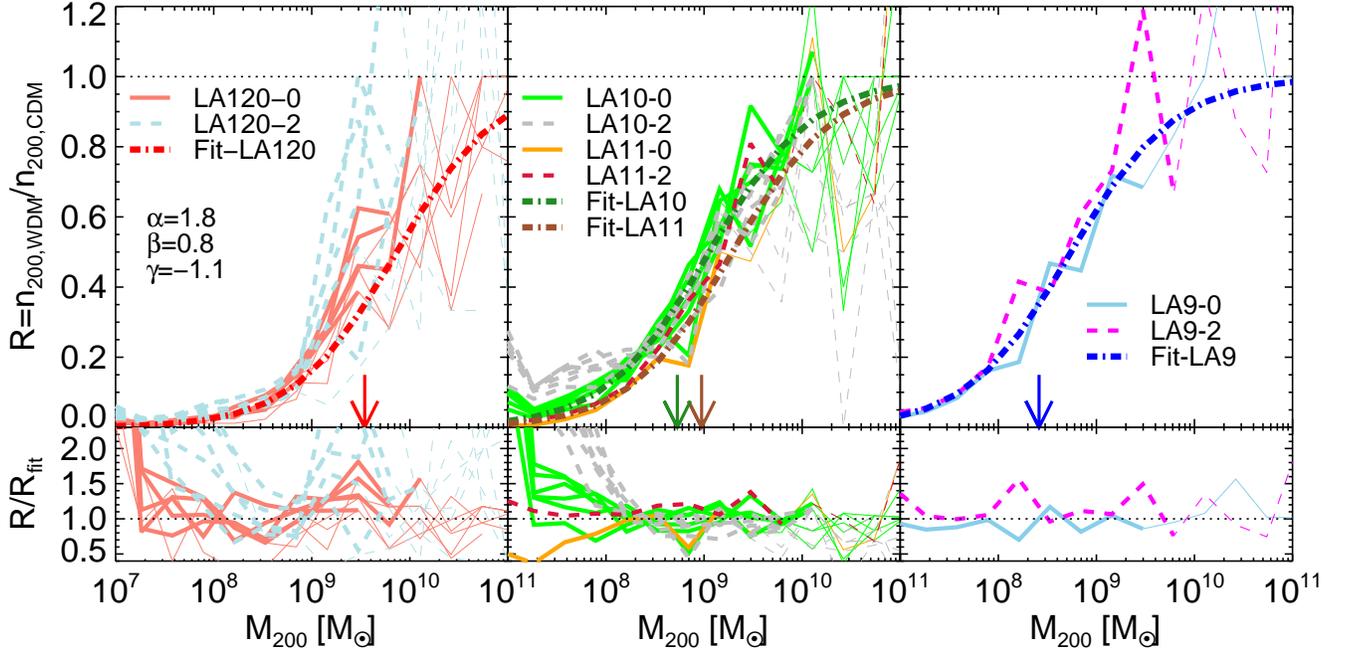}
    \caption{As for Fig.~\ref{fig:m200}, but comparing the $M_{200}$ mass function ratios between the AP $z=0$ simulations (labeled `-0' in the figure legends) and their $z=2$ counterparts (labeled `-2'). The $z=2$ data are shown as dashed lines, with their colours as indicated in the figure legends. }
    \label{fig:m200Z0Z2}
 \end{figure*}

 In this appendix we show the degree to which the mass function ratios differ between $z=0$ and $z=2$. We use the AP simulations, for which we have data at $z=2$, and make the same selection as for the $z=0$ data, while keeping a 3~cMpc co-moving aperture around the center of the zoom region. We plot these data together with the $z=0$ results in Fig.~\ref{fig:m200Z0Z2} for the $M_{200}$ definition of halo mass. We do not use these AP $z=2$ data for the purpose of deriving the fitting parameters, either in this appendix or in our main results.

\begin{figure*}
    \centering
    \includegraphics[scale=0.7]{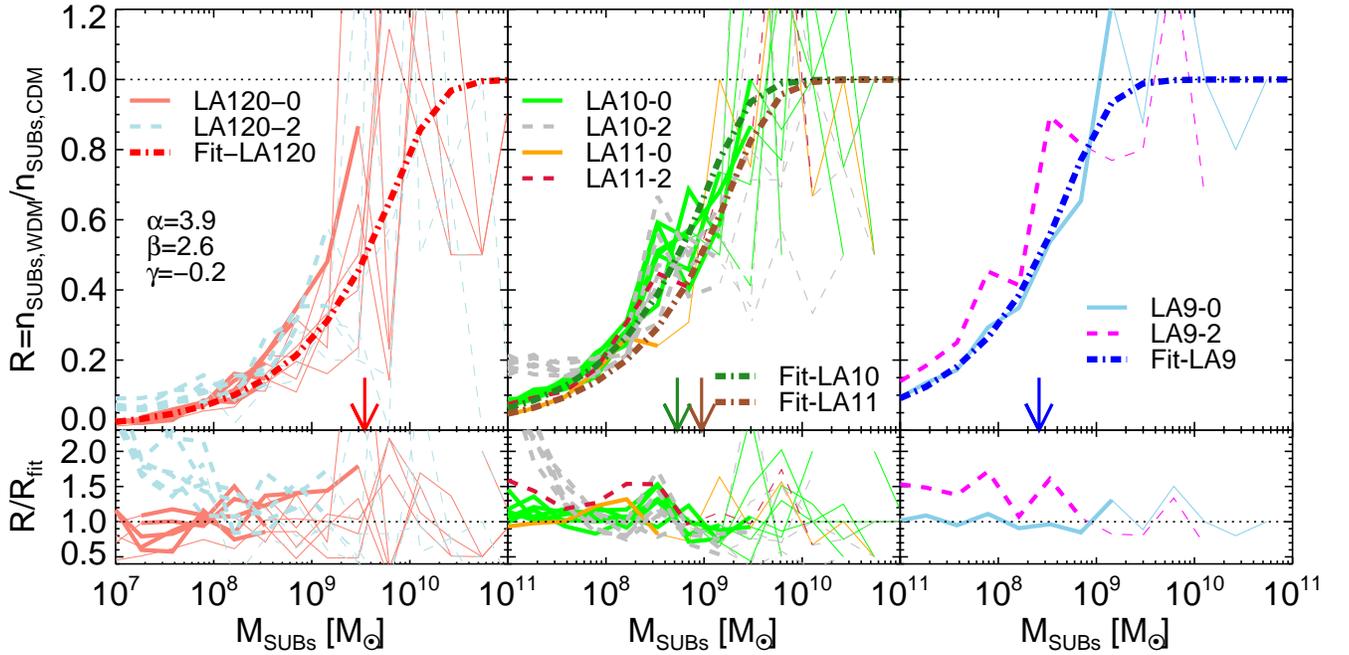}
    \caption{As for Fig.~\ref{fig:m200Z0Z2}, but for subhaloes using the {\sc subfind} bound mass, $M_\mathrm{SUBs}$. }
    \label{fig:msubsZ0Z2}
 \end{figure*}

 We find that the $z=2$ mass function ratios are largely well described by the fitting function derived using the other datasets, and the fit is of roughly the same $\sim10$~per~cent accuracy as for the $z=0$ results for $M_{200}>10^{8}\Msun$. The most notable exceptions are LA120 and LA10, where the fit frequently underestimates the number of $z=2$ haloes, which may be explained in part by poor resolution. We repeat this exercise for the subhalo mass functions in Fig.~\ref{fig:msubsZ0Z2}.
 
The fit systematically underestimates the abundance of $z=2$ AP haloes around $10^{8}\Msun$ at the tens of per~cent level. This result is consistent with the results from the $z=2$ L25-LA120 run, and again likely shows that the suppression of the WDM halo mass function relative to the CDM equivalent becomes stronger towards lower redshift; we expect this phenomenon is due to the correlation between the halo mass -- at fixed redshift -- and the delay in collapse time in WDM simulations \citep{Lovell19b}. The precise reason for why this effect is stronger for subhalo $M_\mathrm{SUBs}$ than for isolated halo $M_{200}$ will be a matter for further study.




\end{document}